\documentclass[12pt]{iopart}

\expandafter\let\csname equation*\endcsname\relax
\expandafter\let\csname endequation*\endcsname\relax
\usepackage{amsmath}

\usepackage{iopams}
\usepackage[square,sort,compress,numbers]{natbib}
\usepackage[pdfusetitle]{hyperref}
\usepackage[dvipsnames, table]{xcolor}
\usepackage[final]{graphicx}
\usepackage{bm, array, multirow}
\usepackage{booktabs}
\usepackage{nicematrix}
\usepackage{enumerate, enumitem}
\usepackage[normalem]{ulem}
\usepackage[capitalize]{cleveref}

\graphicspath{{figs/paper2/}}

\newcommand{\postime}{(\bm{r},t)}
\newcommand{\Rl}{R_{\infty}}
\newcommand{\kl}{k_{\infty}}
\newcommand{\alphac}{\alpha_{c}}
\newcommand{\alphax}{\alpha_{\times}}
\newcommand{\dis}{\Lambda}

\newcommand{\movie}[1]{(See Movie #1)}

\begin{document}

\title{Defect interactions in the non-reciprocal Cahn-Hilliard model}
\author{Navdeep Rana}
\address{Max Planck Institute for Dynamics and Self-Organization (MPI-DS), D-37077 G\"ottingen, Germany}

\author{Ramin Golestanian}
\ead{ramin.golestanian@ds.mpg.de}
\address{Max Planck Institute for Dynamics and Self-Organization (MPI-DS), D-37077 G\"ottingen, Germany}
\address{Rudolf Peierls Centre for Theoretical Physics, University of Oxford, Oxford OX1 3PU, United Kingdom}
\begin{abstract}
    We present a computational study of the pairwise interactions between defects in the recently introduced non-reciprocal Cahn-Hilliard model. The evolution of a defect pair exhibits dependence upon their corresponding topological charges, initial separation, and the non-reciprocity coupling constant $\alpha$. We find that the stability of isolated topologically neutral targets significantly affects the pairwise defect interactions. At large separations, defect interactions are negligible and a defect pair is stable. When positioned in relatively close proximity, a pair of oppositely charged spirals or targets merge to form a single target. At low $\alpha$, like-charged spirals form rotating bound pairs, which are however torn apart by spontaneously formed targets at high $\alpha$. Similar preference for charged or neutral solutions is also seen for a spiral target pair where the spiral dominates at low $\alpha$, but concedes to the target at large $\alpha$. Our work sheds light on the complex phenomenology of non-reciprocal active matter systems when their collective dynamics involves topological defects.
\end{abstract}

\maketitle

\section{\label{sec:introduction} Introduction}

In the world of living and active matter, non-reciprocal interactions are the norm rather than the exception \cite{soto2014, agudo-canalejo2019, saha2020, you2020, saha2022, fruchart2021}. Interactions among active individuals, biological or synthetic, can arise from a variety of complex mechanisms, for example, chemical \cite{soto2014, agudo-canalejo2019}, visual \cite{ballerini2008}, social \cite{helbing1995, helbing2000, rio2018}, programmable \cite{fruchart2021,osat2023}, and wake-mediated \cite{ivlev2015}. The effective breaking of action-reaction symmetry gives rise to interesting phenomenon, otherwise absent at thermal equilibrium, such as, self-propulsion in chemically active mixtures, both at micro and macroscopic scales \cite{soto2014, agudo-canalejo2019}, and buckling instabilities in polar flocks \cite{dadhichi2020}. In the non-reciprocal Cahn-Hilliard model (NRCH) \cite{saha2020,you2020,saha2022,rana2023}, parity and time-reversal (PT) symmetries break spontaneously. A variety of non-equilibrium features emerge, for example formation of travelling density bands \cite{saha2020, you2020}, suppression of coarsening \cite{frohoff-hulsmann2021, saha2020}, and localized states \cite{frohoff-hulsmann2021a}. A variant of the NRCH model with non-linear non-reciprocal interactions exhibits chaotic steady states where PT symmetry is locally restored in fluctuating domains \cite{saha2022}. Non-reciprocal interactions thus provide a generic mechanism that gives rise to wave propagation and the emergence of global polar order in active systems. NRCH model also emerges as a universal amplitude equation from the onset of a conserved-Hopf instability, which occurs in systems with two conservation laws \cite{frohoff-hulsmann2023}. Moreover, a systematic coarse-graining of a microscopic model of phoretically active Janus colloids \cite{Golestanian2019phoretic} also results in a similar phenomenological description \cite{Tucci2024}.

\begin{figure*}
    \centering{\includegraphics[width=0.93\linewidth]{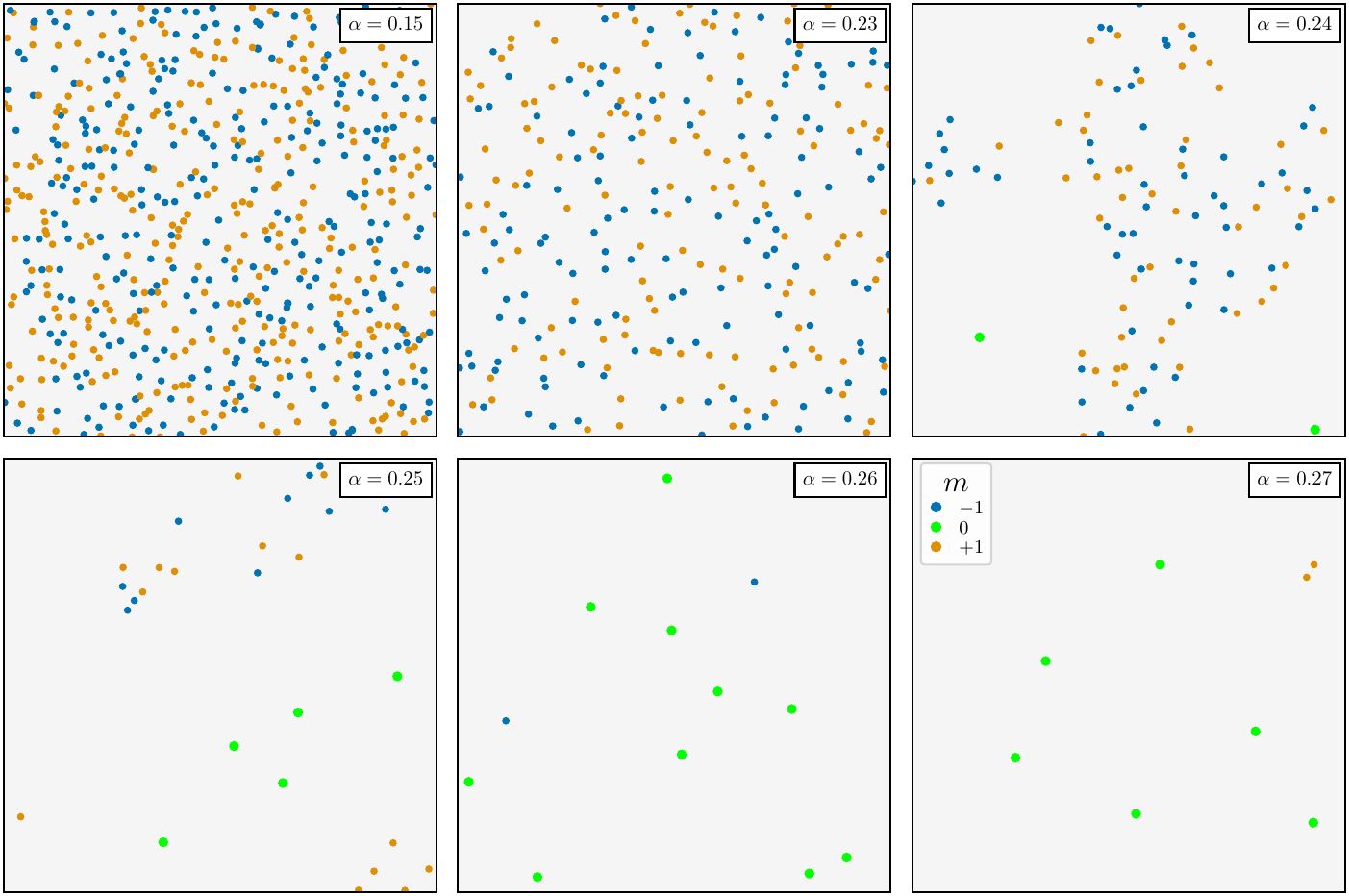}}
    \caption{\label{fig:defect_network_scatter}
        Distribution of spirals and targets identified by the locations of their cores for various values of $\alpha$ and $L = 6400$. At small $\alpha$, the defect networks are composed of isolated and bound pair of spirals. As we increase $\alpha$, targets appear as well and are the dominant population of defects near $\alphac$. Average inter-defect separation increases with $\alpha$ and targets show remarkably higher inter-defect separation as compared to the spirals.
    }
\end{figure*}

Topological defects play an important role in determining the dynamics of systems with broken symmetries \cite{Nelson2002defects}, and have been extensively studied in the context of both equilibrium and non-equilibrium systems \cite{chaikin1995, chandrasekhar1992, romano2023, romano2024}, as well as in quantum gases \cite{HadzibabicNature2006} and cosmology~\cite{Hindmarsh_1995}. Defect unbinding drives the Berezinksii-Kosterlitz-Thouless transition at equilibrium \cite{Kosterlitz1973JPC,chaikin1995}, and an analogous transition in active nematics \cite{shankar2018}, as well as coarsening dynamics of mixtures~\cite{bray2002} and collective properties of type-II superconductors~\cite{Abrikosov2004RMP}. In models of ferromagnets \cite{yurke1993, qian2003, bray2002}, and Toner-Tu model of flocking \cite{rana2020, rana2022}, coarsening proceeds via continuous merging of defects that exhibit (marginally) long-range interactions. Wet suspensions of polar active matter show spectacular defect-mediated flows, arising from the instabilities of the ordered states also known as active turbulence \cite{simha2002,dombrowski2004,ramaswamy2010,uchida2010,uchida2010a,wensink2012,chatterjee2021,rana2024}. Furthermore, topological defects have been shown to play a key role in a wide range of phenomena in active matter \cite{uchida2010,wensink2012,SanchezNature2012,Giomi2013,Thampi2013,martinez-prat2021,Kawaguchi2017nature,Saw2017Nature,MaroudasSacks2020,CopenhagenNatPhys2021}.

There have been extensive studies in which defects have been used as fundamental singularities of the order parameter and effective theories have been developed in terms of the dynamics of the defects \cite{Dafermos1970,IMURA1973403,Eshelby1980PhilMagA,AmbegaokarPRB1980,Dubois-violette,Kawasaki1983linedef,Kawasaki1984Progr,KAWASAKI1984319,BODENSCHATZ1988PhysD,NEU1990PhysD,Pismen1990PRA,Rubinstein1991,RodriguezPRA1991,Pismen1991PhysicaD,DennistonPRB1996,Pleiner1988PRA,Semenov_1999EPL,NajafiEPJB2003,Radzihovsky2015PRL},
including recent related developments in active matter \cite{Karsten2004,PismenPRE2013,TangSM2019,CortesePRE2018,ShankarPRL2018,Vafadefects2020,ZhangPRE2020,AnghelutaNJP2021,VafaSoftMatt2022}. In the case of the complex Ginzburg-Landau (CGL) equation, defect cores have been shown to be screened from outside perturbations, and therefore, defect-interactions become short-ranged; the interactions are independent of charge but can be attractive or repulsive depending on the parameters \cite{aranson1991, aranson1993, aranson2002}.

In \citet{rana2023}, we have studied the defect solutions of the NRCH model, and shown that it admits two types of defects: spirals, which have a unit magnitude topological charge, and topologically neutral targets. For a given strength of non-reciprocity coupling $\alpha$, defect solutions with a unique asymptotic wave number, $\kl \propto \sqrt{\alpha}$ and amplitude $\Rl = \sqrt{1-\kl^2}$ are selected \cite{rana2023}. Wavenumber selection and the Eckhaus instability \cite{rana2023, aranson2002} set a threshold of $\alphax$ above which defect solutions cease to exist. However, our large-scale numerical simulations, which start from random disordered states, reveal a disorder-order transition at $\alphac \ll \alphax$. Below $\alphac$, a disordered state evolves into a quasi-stationary network of defects with vanishing global polar order. Above $\alphac$, we find noisy travelling waves with long-lived fluctuations. The topological composition of the defect networks also changes with $\alpha$. For $\alpha \ll \alphac$, spirals are exclusively selected, whereas for $\alpha \lesssim \alphac$ we primarily observe target networks (see \cref{fig:defect_network_scatter}).

In this paper, we present a detailed numerical investigation of the defect interactions in the NRCH model. In \cref{sec:model}, we introduce our model and provide a brief summary of the properties of isolated defect solutions, previously reported in \cite{rana2023}. In \cref{sec:networks}, we describe the properties of the quasi-stationary defect networks. To explain the features of defect networks, we study interactions between two defects with different relative charges in \cref{sec:interactions}. We find that the stability of the targets leads to fundamentally new kinds of pairwise defect interactions, which are contrasted from interactions in non-equilibrium systems with non-conserved order parameter \cite{aranson2002}. In particular, we observe annihilation of oppositely charged spirals, which leads to formation of neutral targets, and spontaneous creation of targets in between a pair of like-charged spirals at large $\alpha$ \movie{1 and 2}. We conclude the paper with a discussion and possible future directions in \cref{sec:discussion}.

\section{\label{sec:model} Model and Methods}

\emph{Model} -- We consider two conserved scalar fields $\phi_{1}\postime$ and $\phi_{2}\postime$ with non-reciprocal interactions. The complex scalar order parameter $\phi = \phi_1 + i \phi_2$ obeys the following non-dimensional equation
\begin{equation}\label{eq:nrch}\begin{aligned}
    \partial_t{\phi} &= \nabla^2\left[(-1 + i \alpha)\phi + |\phi|^2\phi - \nabla^2 \phi\right],
\end{aligned}\end{equation}
where $\alpha$ measures the strength of the non-reciprocal interactions among $\phi_1$ and $\phi_2$. A detailed description of the model is available in \cite{rana2023, saha2020}. NRCH model \eqref{eq:nrch} admits a variety of novel non-equilibrium states as has been shown recently \cite{saha2020, you2020, saha2022,frohoff-hulsmann2021,frohoff-hulsmann2021a}. Here, we study the properties of the defect networks emerging from the time evolution of an initial disordered state as we have reported recently in \citet{rana2023}.

\emph{Single defect solution} -- A single defect solution of \eqref{eq:nrch} is of the form
\begin{equation}\label{eq:defect}\begin{aligned}
    \phi(\bm{r},t) = R(r) e^{i\left[ m \theta + Z(r) - \omega t\right]},
\end{aligned}\end{equation}
where $r$ (as measured from the defect core) and $\theta$ are the polar coordinates, $R(r)$ is the amplitude, $Z(r)$ is the phase, and $m$ is the topological charge. An isolated spiral core $(m=\pm 1)$ is singular and stationary, thus $R(r)$ vanishes at the origin and is independent of time. On the other hand, a target $(m=0)$ is topologically neutral and its amplitude remains finite at the core while oscillating slowly \cite{rana2023, hagan1982, aranson2002, hendrey2000}.

Irrespective of the value of $m$, defects emanate radially outward travelling waves, and far away from the defect core $(r \gg 1)$ the wave front approaches a plane wave, i.e. $k(r) \equiv d Z(r)/dr \rightarrow \kl$, and $R(r) \rightarrow \Rl = \sqrt{1 - \kl^2}$ \cite{rana2023}. The equivalence of the charged and neutral defects in the context of the generated travelling waves becomes clear by looking at the polar order parameter
\begin{equation}\label{eq:polar}\begin{aligned}
    \bm{J}\postime \equiv \frac{1}{2 i}\left(\phi^{*} \nabla \phi - \phi\nabla\phi^{*}\right).
\end{aligned}\end{equation}
For monochromatic travelling waves, $\bm{J}$ is parallel to the wave vector of the wave. On the other hand, $\bm{J} = R(r)^2 \left(k(r) \hat{r} + \frac{m}{r} \hat{\theta}\right)$ for the spirals and targets. Independently of the value of $m$, $\bm{J}$ has a unit positive topological singularity at the defect core. Far away from the core, $\bm{J} \sim \Rl^2 \kl \hat{r}$ is purely radial. Since $\bm{J}$ is isotropic, the average polar order $\bar{J} \equiv \left|\left<\bm{\widehat{J}}\postime\right>\right| = 0$ for isolated defects. Consequently, disordered defect networks have vanishing average polar order.

\emph{Numerical Methods} -- We use a GPU-accelerated pseudo-spectral algorithm to numerically integrate the equation of motion \eqref{eq:nrch} on a two-dimensional periodic box with sides $(L_x, L_y)$ discretized over $(N_x, N_y)$ points. For time marching, we use a second-order exponential time-differencing scheme \cite{cox2002}. For simulations that start from random initial conditions, we use the same setup as used in Ref. \cite{rana2023} and set $L_x = L_y = L$ and $N_x = N_y = N$. The setup for the defect interaction simulations follows \cite{aranson1993} and is described in \cref{sec:interactions}.

\begin{figure}[t]
    \centering{\includegraphics[width=0.5\linewidth]{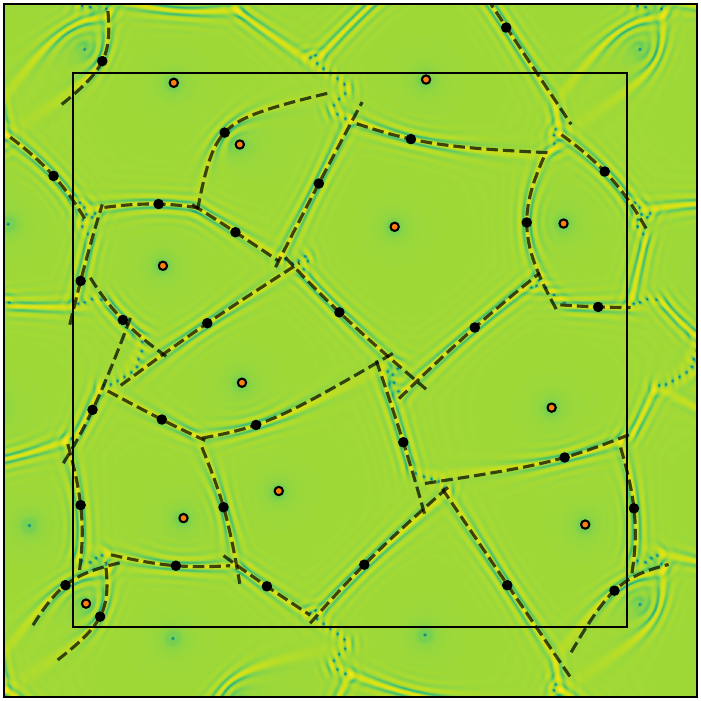}}
    \caption{\label{fig:hyperbole}
        Hyperbole construction \cite{bohr1997} for a particular realisation of a defect network for $\alpha=0.18$. We extend the periodic domain (shown by solid black square) on all four sides to generate the construction. Constructed hyperbole segments are shown by dashed black lines. Orange circles mark the location of the defect cores that are the foci of the hyperbole (all spirals for this particular $\alpha$), whereas black circles mark the points on the disinclination lines that were used to measure the distance from the foci.
    }
\end{figure}

\emph{Finding the defect cores} -- To identify the spiral and target cores, we use a topological charge counting algorithm \cite{berg1981} adopted for two dimensions. While the algorithm locates the spirals cores in a straightforward manner, it cannot find the location of the topologically neutral targets. In addition, the algorithm also gives the undesirable topological zeroes present at the intersection of the disinclination lines (See \cref{fig:hyperbole}, for example). Both of these issues are easily resolved by exploiting the properties of the polar order parameter $\bm{J}\postime$ \eqref{eq:polar}. To find the defect cores of $\phi$, we first find the topological zeros of the polar current $\bm{J}$ carrying a positive unit charge. We then calculate the topological charge of $\phi$ to classify them into spirals and targets. To eliminate the defects on the disinclination lines, we introduce the quantity $Q = \frac{1}{A_\Omega}\int_\Omega \bm{J} \cdot \bm{\Delta r} d\Omega$, where $\bm{\Delta r}$ is measured from the defect core, and $\Omega$ is a small domain around the defect with area $A_\Omega$. As $\bm{J}\sim \widehat{r}$, $Q$ is large and positive for isolated spirals and targets. On the other hand, $\bm{J}$ changes with orientation for the defects on the disinclination lines and the absolute value of $Q$ is small. A suitable threshold of $Q$, which we choose heuristically, then filters out the defects on the disinclination lines.

\section{\label{sec:networks} Disordered defect networks}

For non-reciprocity coupling smaller than the critical threshold $\alphac$, an initially disordered configuration evolves into a quasi-stationary defect network. The defect network emerges after an initial transient period in which numerous newly-born defects move around and merge to form a stable configuration \cite{rana2023}. In \cref{fig:defect_network_scatter}, we plot the location of the defects in the defect networks for various values of $\alpha$. The defect networks are composed of isolated spiral and target cores separated from each other by distances larger than the core radius. At low values of $\alpha$, we observe isolated spirals and a handful of bound pairs of like-charged spirals that revolve around a common centre. As we increase $\alpha$, targets emerge and dominate the system for $\alpha$ just below $\alphac$. Inter-defect separation increases with $\alpha$ albeit without showing a clear trend. However, targets show a strikingly higher inter-defect separation compared to the spirals. The spatial distribution of the defects depends highly on the initial conditions. Consequently the defect density differs significantly across various realizations of the initial disordered state. For $\alpha < \alpha_c$, our tests show that a randomly chosen defect configuration respecting the typical inter-defect separation and defect density always forms a stable defect network.

Disinclination lines are formed where the waves generated by two or more defects meet. The shape of the disinclination lines is easily determined by phase matching of the nearby spiral/target cores \cite{bohr1997,aranson2002}. As shown in \cref{fig:hyperbole}, they are well approximated as segments of hyperbole whose foci are the two nearest defects, i.e., $r_1 - r_2 = c$, where $c$ is a constant and $r_1$ and $r_2$ are the distances from the two nearest defects (with $r_1 < r_2$). Additional topological zeros of the $\phi$-field are found where multiple disinclination lines meet. These defects are separated from each other by distances comparable to the core radius $\ell$ and do not play a major role in determining the features of the defect network.

Disordered networks are effectively frozen in time. The dynamics of the spiral and target cores is limited to the emanated travelling waves. While doing so, the arms of the isolated spirals rotate with the frequency $\omega$, the targets pulsate, and the spiral cores that form bound pairs orbit around a common centre. Additional defects found on the disinclination lines occasionally undergo periodic dynamic rearrangement as they are unable to settle down into a stable stationary state \cite{rana2023}.

\begin{figure}[t]
    \centering{\includegraphics[width=0.5\linewidth]{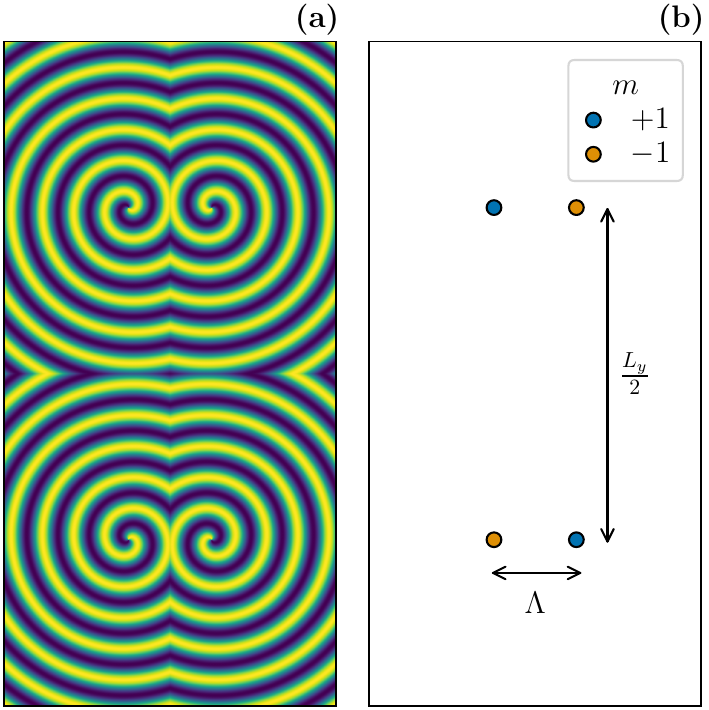}}
    \caption{\label{fig:interaction_setup}
        Simulation setup used to study interactions among defect pairs. (a) Initial condition for like-charged spirals and (b) schematic diagram. There are two defect pairs, one in the lower half of the domain and the other in the upper half of the domain. Defect pairs are far apart from each other ($\frac{L_y}{2} > 2 \dis$), and thus are expected to have negligible interactions between them. Because of the reflection symmetry with respect to the $y = \frac{L_y}{2}$ line, the spirals in upper pair have the opposite charge as compared to the spirals in the lower pair.
    }
\end{figure}

\section{\label{sec:interactions} Defect Interactions}

The quasi-stationary nature of the defect networks hints that the interactions between defects may fall rapidly with increasing separation. Similar behaviour is observed for the CGL equation, where the defect interactions decay exponentially \cite{aranson1991, pismen1992, aranson1993}. Thus we expect that the nearest neighbour interactions are enough to determine the properties of the defect networks. To this end, we numerically study the interactions of a defect-pair with different relative topological charges separated by a varying initial distance $\dis$. For these simulations, we adopt the geometry shown in \cref{fig:interaction_setup}, which has been previously used to study spiral interactions in the CGL equation \cite{aranson1991}. We consider a rectangular box with sides $(L_x, L_y) = (L, 2 L)$ divided into four equal quadrants of size $(L/2, L)$ to approximate no-flux boundary conditions in a periodic domain. A defect is placed in the lower left quadrant at coordinates $\left(\frac{L}{2} - \frac{\dis}{2}, L\right)$. The field is then extended to the upper left quadrant by symmetric reflection along the line $y=L$ resulting in a pair of oppositely charged spirals in the left half of the domain. A further symmetric (anti-symmetric) reflection along $x=L/2$ line gives us a pair of oppositely (like) charged spirals separated by a distance $\dis$ apart (See \cref{fig:interaction_setup}). Unlike the study of \citet{aranson1993}, we perform this symmetrization only at the beginning of the simulation. Since the targets do not break the mirror symmetry, the same procedure is also used to generate target pairs. This procedure leads to two pairs of defects, or a total four defects in the rectangular domain. As $\dis < L$, the dominant interactions are among the horizontal neighbours and the two pairs evolve independently to each other in an identical manner. Thus, in the following discussion, we focus on the dynamics of one pair only. Finally, we note here that we have performed a systematic numerical study of the defect interactions for $\alpha \leq 0.35 < \alphax$. As we move closer to $\alphax \sim 0.58$, numerical perturbations and finite-size effects tend to destabilize the initial defect-pair configurations or the final steady-states.

In the rest of this paper, we will show that defects in the NRCH model show new kind of interactions, largely facilitated by the stability of the topologically neutral targets. As summarized in \cref{tab:defect}, the evolution of the pair depends on the relative topological charges on the defects, the initial separation $\dis$, and the strength of the non-reciprocal interactions $\alpha$. As expected, at large initial separation $\dis$, irrespective of the relative topological charges, interactions are negligible and the defect pair remains stationary or evolve at extremely slow time scales. In what follows, we describe our results for small and intermediate separations for different $\alpha$ values, and use them to explain the properties of the defect networks.

\newcommand{\spc}[1]{\cellcolor{blue!20}#1}
\newcommand{\tgc}[1]{\cellcolor{green!20}#1}
\begin{figure}
    \centering
    \renewcommand{\arraystretch}{1.1}
    \begin{tabular}{wl{2.8cm}wc{2.6cm}wc{2.6cm}wc{2.6cm}wc{2.6cm}}
        \toprule
        \multirow{2}{*}{Defect pair} & \multicolumn{2}{c}{Low $\alpha$} & \multicolumn{2}{c}{High $\alpha$} \\
        \cmidrule(lr){2-3}\cmidrule(lr){4-5}
                              & Small $\dis$           & Large $\dis$            & Small $\dis$    & Large $\dis$\\
                              \midrule
        Targets               & \tgc{Target}        & \tgc{Stable Targets} & \tgc{Target} & \tgc{Stable Targets} \\
        Spiral-Target         & \spc{Spiral}        & \spc{Spiral}         & \spc{Spiral} & \tgc{Target} \\
        Like Spirals          & \spc{Bound Spirals} & \spc{Stable Spirals} & \tgc{Target} & \tgc{Target} \\
        Opposite Spirals      & \tgc{Target}      & \spc{Stable Spirals} & \tgc{Target} & \spc{Stable Spirals} \\
        \bottomrule
    \end{tabular}
    \caption{\label{tab:defect}
        Summary of pairwise defect interactions for different values of the relative charge, initial separation $\dis$, and
        non-reciprocity coupling $\alpha$. For each case, we list the final state.
    }
\end{figure}

\subsection{Like-charged Spirals}
The evolution of a like-charged spiral pair shows two distinct regimes depending upon the strength of the non-reciprocity coupling $\alpha$.

\begin{figure}
    \centering{\includegraphics[width=\linewidth]{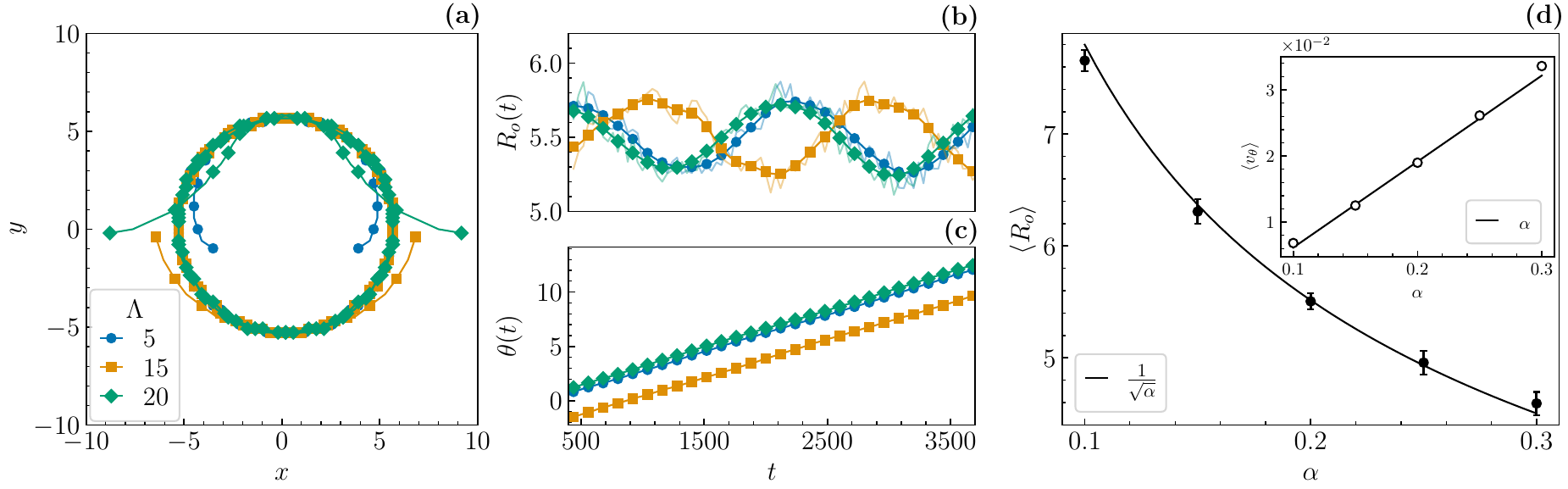}}
    \caption{\label{fig:like_charged_evolution}
        (a) Evolution of two like-charged spirals to a well-defined orbit for $\alpha=0.2$. Irrespective of the initial separation, the spiral pair settles in an orbit with average radius $\left<R_o\right> \sim 5.5$. (b) Orbit radius varies sinusoidally with time. Light lines show the actual orbit over time, and the markers show the smoothed data. (c) Orientation increases linearly with time. (d) Plot of average orbit radius $\left<R_o\right> \sim 5.5$ versus $\alpha$. Error bars indicate root mean-square deviation from the mean. Solid black line shows the fit $1/\sqrt{\alpha}$. (Inset) Average tangential velocity versus $\alpha$; the error bars are smaller than the marker size. Solid black line shows the fit $\alpha$.
    }
\end{figure}

\begin{figure}
    \centering{\includegraphics[width=0.5\linewidth]{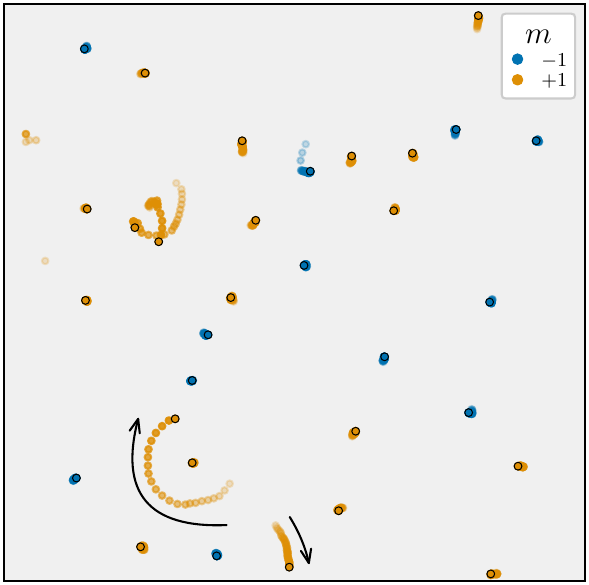}}
    \caption{\label{fig:motion} Motion of spiral cores in a disordered defect network. Time is marked by increasing
        opacity of the markers and the direction of motion is highlighted for two spirals using black arrows. The final
        position of each defect is marked by a black hollow marker. A few spiral cores are easily seen to orbit around
    like charged neighbours. }
\end{figure}

\emph{Small $\alpha$ --} As shown in \cref{fig:like_charged_evolution}, for small $\alpha \lesssim 0.3$ and small initial separations, two like-charged spirals settle in an orbit and revolve around a common centre \movie{1}. These orbiting like-charged spirals can be thought of as a bound-state. The radius of the orbit, $R_o(t)$, and the angular velocity, $v_\theta$, depend on $\alpha$, but are independent of the initial separation (See \cref{fig:like_charged_evolution}(a)). Furthermore, $R_o(t)$ varies sinusoidally with time, although the total variation is small compared to its average value (i.e. $< 10\%$, See \cref{fig:like_charged_evolution}(b)). We find that the average radius decreases as $\left<R_o(t)\right> \sim 1/\sqrt{\alpha}$, and the angular velocity increases linearly with $\alpha$ (See \cref{fig:like_charged_evolution}(c)). As we increase the initial separation $\dis$, the orbits appear to become larger in a quantized manner (data not shown here), although the dynamics is very slow. 
We have also observed a pair of like-charged spirals orbiting around each other in our simulations starting from the disordered states (See Movie 2 in \cite{rana2023}). In addition, we occasionally found a spiral moving on a circular arc (See \cref{fig:motion}), while its neighbours remain stationary. A visual inspection of the spiral and its neighbours reveals that they all carry the same topological charge. The motion of the spirals then arises from considering unbalanced pairwise interactions with its immediate neighbours.

\emph{Large $\alpha$ --} For $\alpha \gtrsim 0.3$, the spirals initially form a bound pair and start to rotate around each other. At later times, however, a target forms spontaneously on the line joining the spiral cores and pushes the spirals outwards \movie{1}. Since periodicity ensures that the net charge over the entire domain is zero, we thus obtain a target defect at the centre of the domain at long times.

\subsection{Oppositely-charged Spirals}
When placed closer than a threshold separation $\dis_{SS}$, a pair of oppositely-charged spirals annihilate to form a single target. This scenario is allowed because targets are stable solutions of the NRCH model \eqref{eq:nrch}, but it has no counterpart in the CGL equation. As shown in \cref{fig:opposite_charged_threshold}, the threshold separation increases with $\alpha$, which again explains the apparent increase in the typical inter-defect separation with $\alpha$. When placed at distances larger than the threshold value, the spiral pair is stable and moves in a direction perpendicular to the line joining their centres. Similar to the oppositely-charged spirals in the CGL equation \cite{aranson1993}, the tangential velocities for the two spirals are parallel, and decrease as the initial separation increases. At large separation distances, the two spirals have negligible interactions \movie{2}.

\subsection{A Target pair}
A pair of targets shows similar features as a pair of oppositely-charged spirals. Below a threshold initial separation, the pair merges to form a single target. Above the separation distance, the pair is stable. However, as they are topologically neutral, the targets in the stable pair remain stationary. Once again, we find that the threshold separation increases with increasing $\alpha$ \movie{3}.

\subsection{A Spiral and a Target}
\begin{figure}
    \centering{\includegraphics[width=0.9\linewidth]{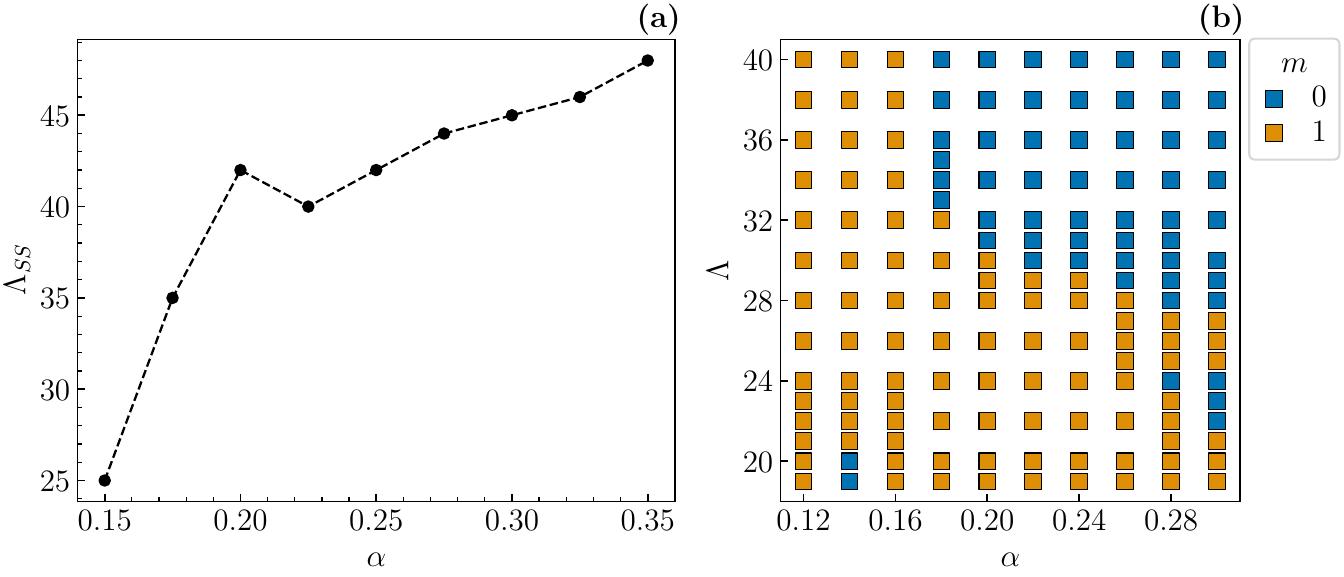}
    \caption{\label{fig:opposite_charged_threshold} (a) Plot of the threshold separation $\dis_{SS}$ for various values of $\alpha$ for a pair of oppositely charged spirals. For $ \dis < \dis_{SS}$, the two spirals merge to form a target, while for $\dis > \dis_{SS}$, the pair is stable. (b) Phase plot showing final state of a spiral-target pair at various values of $\alpha$ and initial separation $\dis$. At low $\alpha$, the pair always evolves into a single spiral. As we increase $\alpha$, beyond a threshold separation distance, the target helps to disintegrate the core of the spiral. $\dis_{ST}$ increases with $\alpha$.}}
\end{figure}

We summarize the fate of a spiral-target pair in the $\alpha-\dis$ stat diagram shown in \cref{fig:opposite_charged_threshold}(b). For $\alpha < 0.18$, irrespective of the initial separation, a spiral-target pair eventually merges to form a single spiral. For $\alpha \geq 0.18$, a threshold of initial separation $\dis_{ST}$ marks the change in the final state. For $\dis < \dis_{ST}$, the pair evolves into a single spiral. Above $\dis_{ST}$, the outward travelling waves emitted from the target tear the spiral core apart pushing it radially outwards. Owing to the periodic boundary conditions, we are left with a single target. The threshold separation $\dis_{ST}$ increases with $\alpha$. We note that there are a few exceptions to this overall consistent behaviour, at small values $\alpha$.

\section{\label{sec:discussion} Discussion}

We have presented here a detailed numerical analysis of the interaction of defects in the NRCH model. While both charged and neutral solutions are allowed for a given $\alpha$, our study reveals that the system prefers to spontaneously break chiral symmetry at small $\alpha$ and select spirals as the defect solutions. On the other hand, at large values of $\alpha$, targets are the dominant defects. The defect solutions of the NRCH model share some features with those of the CGL equation, but exhibit novel types of interactions, facilitated mainly by the stability of the topologically neutral target solutions. Importantly, we find that defect interactions depend on the relative topological charges as well as the strength of non-reciprocity. Furthermore, the time evolution of defect pairs readily explains the dynamics of the observed defect networks. While we have performed large-scale and long-time simulations, it would be interesting to see the behaviour of defect networks at even longer timescales. For example, defect states of the CGL equation evolve extremely slowly \cite{brito2003}, and can show a vortex liquid phase where spirals exhibit normal diffusion or a vortex glass state with intermittent slow relaxation. Whether defect states of the NRCH model show similar features remains to be explored. 

\appendix

\section{\label{sec:movies} Description of the movies}

\begin{itemize}[leftmargin=*]
    \item \texttt{\textbf{movie1\_like\_charged.mkv}} shows the evolution of a like-charged spiral pair configuration for $\alpha = 0.20~\rm{and}~ 0.35$. At $\alpha = 0.2$, the spirals form a bound pair and orbit around a common point with constant angular velocity (See \cref{fig:like_charged_evolution}). At $\alpha=0.35$, a target emerges spontaneously at the centre of the line joining the spiral pair and pushes the spirals away. Since the net charge on the periodic domain is zero, the final state is a target.
    \item \texttt{\textbf{movie2\_opposite\_charged.mkv}} shows the evolution of an oppositely charged pair of spirals for $\dis=30~\rm{and}~50$ at $\alpha=0.2$. For $\dis = 30 < \dis_{SS}$, the spirals merge and form a single target (See \cref{fig:opposite_charged_threshold}(a)). For $\dis = 50 > \dis_{SS}$ the pair remains stable and move in a direction perpendicular to the line joining their centres.
    \item \texttt{\textbf{movie3\_targets.mkv}} shows the time evolution of a target pair for $\dis=25~\rm{and}~50$ at $\alpha=0.2$. The dynamics of the target pair is similar to a pair of oppositely charged spirals. For $\dis = 25$, the target pair merge and form a single target. For $\dis = 50$ the pair remains stable and stationary.
    \item \texttt{\textbf{movie4\_spiral\_and\_target.mkv}} shows the time evolution of a spiral-and-target pair for $\dis=25~\rm{and}~40$ at $\alpha=0.2$, where a threshold separation $\dis_{ST}$ marks the change in behaviour. For $\dis = 25 < \dis_{ST}$, the pair merges to form a single spiral, whereas for $\dis = 40 > \dis_{ST}$ the pair merges to form a single target (See \cref{fig:opposite_charged_threshold}(b)).
\end{itemize}

\bibliography{Bibliography}
\bibliographystyle{unsrtnat}

\end{document}